\documentclass{article}
\usepackage[utf8]{inputenc}
\usepackage{amsmath, amssymb, amsthm, graphicx, url}
\usepackage{multirow}
\usepackage{tabularx} % TODO: Check this is okay.
\usepackage{fullpage} % TODO: Remember to delete.
\usepackage{xcolor} % TODO: Remember to delete.
\usepackage{booktabs}
\usepackage{hyperref}
\usepackage{cite}
\usepackage{enumitem}

\newcommand{\adfa}{A\nobreakdash-DFA}
\newcommand{\ddfa}{D\textsuperscript{2}FA}

\newcommand{\simi}[0]{\textit{sim}}

\newcommand{\algoriginalfull}[0]{\textsc{D\textsuperscript{2}FA}}
\newcommand{\algoriginalsparse}[1]{\textsc{Sparse-D\textsuperscript{2}FA}}

\newcommand{\algrefinedfull}[1]{\textsc{D\textsuperscript{2}FA-{Ld}}}
\newcommand{\algrefinedsparse}[1]{\textsc{Sparse-D\textsuperscript{2}FA-{Ld}}}

\newcommand{\algcutfull}[1]{\textsc{D\textsuperscript{2}FA-{Ld-Cut}}}
\newcommand{\algcutsparse}[1]{\textsc{Sparse-D\textsuperscript{2}FA-{Ld-Cut}}}
\newcommand{\algcutboth}[1]{(\textsc{Sparse-})\algcutfull{}}

\newcommand{\algadfafull}[1]{\textsc{D\textsuperscript{2}FA-{Md}}}
\newcommand{\algadfasparse}[1]{\textsc{Sparse-D\textsuperscript{2}FA-{Md}}}

\title{Fast Practical Compression of Deterministic Finite Automata\footnote{An extended abstract appeared at the \emph{50th International Conference on Current Trends in Theory and Practice of Computer Science}~\cite{BGP2025}.}}
\author{Philip Bille \\ \texttt{phbi@dtu.dk} \and Inge Li G{\o}rtz \\ \texttt{inge@dtu.dk} \and Max Rish{\o}j Pedersen \\ \texttt{mhrpe@dtu.dk}}
\date{}

\begin{document}
\maketitle

\begin{abstract}      
    We revisit the popular \emph{delayed deterministic finite automaton} (\ddfa{}) compression algorithm introduced by Kumar~et~al.~[SIGCOMM 2006] for compressing deterministic finite automata (DFAs) used in intrusion detection systems. This compression scheme exploits similarities in the outgoing sets of transitions among states to achieve strong compression while maintaining high throughput for matching. 
    
    The \ddfa{} algorithm and later variants of it, unfortunately, require at least quadratic compression time since they compare all pairs of states to compute an optimal compression. This is too slow and, in some cases, even infeasible for collections of regular expression in modern intrusion detection systems that produce DFAs of millions of states. 

    Our main result is a simple, general framework for constructing \ddfa{} based on locality-sensitive hashing that constructs an approximation of the optimal \ddfa{} in near-linear time. We apply our approach to the original \ddfa{} compression algorithm and two important variants, and we experimentally evaluate our algorithms on DFAs from widely used modern intrusion detection systems. Overall, our new algorithms compress up to an order of magnitude faster than existing solutions with either no or little loss of compression size. Consequently, our algorithms are significantly more scalable and can handle larger collections of regular expressions than previous solutions.       
\end{abstract}

\section{Introduction}\label{sec:intro}
Signature-based deep packet inspection is a key component of modern intrusion detection and prevention systems. The basic idea is to maintain a collection of regular expressions, called \emph{signatures}, that correspond to malicious content, violations of security policies, etc., and then match the collection of signatures against the input. In the typical scenario of high-throughput network traffic, the matching must be fast enough to process the input at the network's speed. 

A natural approach to solve this is to construct a \emph{deterministic finite automaton} (DFA) of the collection of signatures and then simulate it on the input. The DFA matches each character of the input with a single constant time state transition and requires only a single memory access. Unfortunately, DFAs for the collections of signatures used in modern intrusion detection systems are prohibitively large and not feasible for practical implementation~\cite{YCDL+2006, BC2007}. To overcome this space issue while still maintaining fast matching, significant work has been done on compressing DFAs~\cite{BC2008, BTC2006, KSE2008, TSCV2004, AFSK+2012, BC2007a, KDYP+2006, FGPV+2008, FPGP+2011, AFSK+2015, QWFX+2011, SLHW2017, MPNT+2010, PLT2014, LT2014, BC2013, KTW2006, BC2007b, LSLH+2017, MMK2018, GWXC+2023}.

%\paragraph{Delayed Deterministic Finite Automata}
In this paper, we revisit the elegant and powerful \emph{delayed DFA} compression technique introduced by Kumar et al.~\cite{KDYP+2006} and applied in many subsequent solutions~\cite{MPNT+2010, PLT2014, LT2014, BC2013, KTW2006, BC2007b, LSLH+2017, MMK2018, MKMK2018, GWXC+2023}. The basic observation is that many states in the DFAs for real-world regular expression collections have similar sets of outgoing transitions, and we can take advantage of this to reduce the space significantly. Specifically, if two states $s$ and $s'$ share many such transitions, we can replace these in $s$ with a special \emph{default transition} to the other state $s'$. 

Kumar et al.~\cite{KDYP+2006} proposed an algorithm to compute an optimal set of default transitions to compress any DFA. The key idea is to compute the similarity of all pairs of states, i.e., the number of shared outgoing transitions. We store these in a complete graph, called the \emph{space reduction graph} (SRG), on the states with edges weighted by similarity. Finally, we compute a maximum spanning tree on the SRG and use each edge in the tree as a default transition leading to a compressed version of the original DFA called the \emph{delayed deterministic finite automaton} (\ddfa).

To implement matching, we traverse the \ddfa{} similar to the Aho-Corasick algorithm for multi-string matching~\cite{AC1975}. When we are at a state $s$ and want to match a character $\alpha$, we first inspect the outgoing transitions at $s$ for a match of $\alpha$. If we find a match, we continue to that state and process the next character in the input, and if not, we follow the default transition to state $s'$ and repeat the process from $s'$ with character $\alpha$. 

Compared to a standard DFA solution, Kumar et al.~\cite{KDYP+2006} showed that the \ddfa{} dramatically reduces space by more than 90\% on real-world collections of regular expressions while still achieving fast matching performance.   

%\paragraph{Longest Delay and Matching Delay}  
Processing a character \ddfa{} during matching may require following multiple default transitions, thus incurring (as the name suggests) a \emph{delay}. Minimizing the delay is essential in high-throughput applications, and two important variants of \ddfa{}s that address this problem have been proposed. Kumar et al.~\cite{KDYP+2006} gave a modified \ddfa{} construction that, given an integer parameter $L$, limits the maximum path of default transitions to $L$. We refer to this as the \emph{longest delay} variant of the problem. Alternatively, Becchi and Crowley~\cite{BC2007b, BC2013} gave a modified \ddfa{} construction that limits the maximum total number of default transitions traversed on any input string $S$ by $|S|$. We refer to this as the \emph{matching delay} variant. 

%\paragraph{Compression Time}
The main bottleneck in the above algorithms is computing the similarity of all pairs of states to construct the SRG. If the input DFA contains $n$ states, this requires at least $\Omega(n^2)$ time and space. This is too slow and, in some cases, even infeasible for collections of regular expression in modern intrusion detection systems that produce DFAs of millions of states. %Thus, a natural question is if we can significantly improve the compression time. 

\subsection{Contributions}
We present a simple, general framework for fast compression of DFAs with default transitions based on locality-sensitive hashing. We apply our approach to general \ddfa{} compression and the longest delay and matching delay variant and experimentally evaluate our algorithms on collections of regular expressions used in the popular Snort~\cite{Roesch1999}, Zeek~\cite{Paxson1999}, and Suricata~\cite{suricata2010} intrusion detection systems (see also \url{www.snort.org}, \url{zeek.org}, and \url{suricata.io}). Overall, we obtain new algorithms that compress up to an order of magnitude faster than existing solutions with either no or little loss of compression size. Consequently, our algorithms are significantly more scalable and can handle larger collections of regular expressions than previous solutions.

Technically, our main idea is to use locality-sensitive hashing to identify approximately similar states quickly. We then add edges between these states weighted by their similarity, producing the \emph{sparse space reduction graph} (SSRG). We then compute the maximum spanning tree on the SSRG and then the \ddfa{}. The SSRG contains significantly fewer edges, leading to a significant improvement in compression time. While the SSRG approximates the SRG by discarding edges, which may lead to worse compression size, we observe that this loss is negligible experimentally.  For the longest delay variant, the previous solution by Kumar et al.~\cite{KDYP+2006} also uses a costly heuristic to maintain a bounded diameter maximum spanning forest. If we directly apply our sparsification technique, this heuristic dominates the running time, and we do not experimentally observe a significant speed-up in compression time. Instead, we develop an efficient alternative heuristic that first constructs a maximum spanning tree and then cuts edges to achieve the desired maximum spanning forest with bounded diameter. We show that combining our sparsification with the new heuristic leads to improvements in compression time similar to our other variants with little or no loss of compression size.

\subsection{Related Work}\label{sec:related_work}
Substantial work has been done on compression DFAs. For an overview, see surveys~\cite{XCSY+2016, SS2017}. A popular approach is compressing the set of transitions~\cite{MPNT+2010, TSCV2004, AFSK+2012, BC2007a, KDYP+2006, KTW2006, BC2007b, FGPV+2008, FPGP+2011, BC2013, PLT2014, LT2014, AFSK+2015,QWFX+2011, LSLH+2017, MMK2018, SLHW2017}. This approach includes the popular \ddfa{} algorithm we focus on in this paper. Another approach is to compress the alphabet to reduce the size of the transition table~\cite{BC2008, BTC2006, KSE2008, BC2013, TJDS+2017}. The main idea is that, if some sets of characters (almost) always cause the same transitions throughout the DFA, they can be replaced by a single character~\cite{BC2008, BTC2006, BC2013, KSE2008}. Alternatively, we can also compress the alphabet by replacing infrequent characters with sequences of frequent characters~\cite{TJDS+2017}. Finally, we can also compress the set of states as proposed by Becchi and Cambadi~\cite{BC2007a}. They showed how states with similar sets of outgoing transitions could be merged into one, thus compressing the set of states.  

Using locality-sensitive hashing for fast compression of collections of sets has been used widely in many other contexts~\cite{DI2003, OMST2002, DAS2010, KDLT2004, PWZ2011, SHWH2012, KH2015, XJFH2011, BGPT2023}. Our work naturally extends this work to fast DFA compression.

\section{Preliminaries}\label{sec:prelims}
\paragraph{Graphs} A \emph{graph} $G = (V, E)$ is a set of \emph{nodes} $V$ (also called vertices) and a set of \emph{edges} $E : V \times V$ between nodes.
We call the two nodes of an edge its \emph{endpoints}.
If the edges have direction, i.e., $(u, v) \neq (v, u)$, then the graph is \emph{directed}, otherwise it is \emph{undirected}.
Edges can have an associated \emph{weight}, in which case the graph is \emph{weighted}.
An edge can also have an associated \emph{label}, in which case the graph is \emph{labeled}.
We denote an edge from $u$ to $v$ with label $c$ as ${(u, v)}_{c}$ and say $(u, v)$ is \emph{$c$-labeled}.
A path of length $k$ between two nodes $u_{0}$ and $u_{k}$ is a sequence of nodes $u_{0}, \ldots, u_{k}$ such that $(u_{i}, u_{i+1}) \in E$ for $0 \leq i < k$.
It can also be viewed as the sequence of edges $(u_{0}, u_{1}), \ldots, (u_{k-1}, u_{k})$.
If $u_{0} = u_{k}$ then $p$ is a \emph{cycle}.
A set of nodes where every pair of nodes has a path between them is called a \emph{connected component}.
If a graph has only one connected component, it is \emph{connected}, otherwise it is \emph{disconnected}.

A \emph{tree} is a connected graph that contains no cycles.
Due to this property, a tree with $n$ nodes has $n-1$ edges.
A graph consisting of several trees is a \emph{forest}. A node with only one incident edge is called a \emph{leaf}. For a node $v$ in a tree, its \textit{radius} is the length of the longest path from $v$ to a leaf. The \emph{diameter} of a tree is the length of the longest path between any two nodes in the tree. A \emph{spanning tree} is a subgraph containing all the nodes and is a tree. If the graph is weighted, the \emph{value} of a spanning tree is the sum of the weights of the edges in the tree. A \emph{maximum spanning tree} (MST) is a spanning tree of maximum value.

\paragraph{Deterministic Finite Automata}
A \emph{deterministic finite automaton} (DFA) is a 5-tuple $D = (Q, \Sigma, \delta, q_0, A)$ where $Q$ is a set of states, $\Sigma$ is an alphabet, $\delta : Q \times \Sigma \rightarrow Q$ is a transition function, $q_0 \in Q$ is the initial state and $A \subseteq Q$ is a set of accepting states.
We let $n = |Q|$ denote the number of states.
A DFA can be thought of as a \emph{labeled directed graph} where $Q$ is the set of nodes and each transition $\delta(u, c) = v$ is a labeled, directed edge~${(u, v)}_{c}$.
See Figure~\ref{fig:example}~(A) for an example.
For simplicity, we assume every state has exactly one labeled transition for each character in the alphabet, i.e., $\delta$ is \emph{total}, as in previous work.

Given a string $S$ and a path $p$ in $D$ we say that $p$ matches $S$ if the concatenation of the labels of $p$ equals $S$. A path that starts in $q_{0}$ and ends in $A$ is \emph{accepting} and $D$ \emph{accepts} a string $S$ if there exists an accepting path that matches $S$. The \emph{language} of $D$ is the set of strings it accepts.

\paragraph{Locality-Sensitive Hashing} A family of hash functions is \emph{locality-sensitive}, for some similarity measure, if the probability of two objects hashing to the same value is \emph{high} (lower-bounded for some parameter) when they are \emph{similar} (similarity above some threshold) and, conversely, \emph{low} when they are \emph{dissimilar} (see e.g.~\cite{PIM2012} for formal details). 
There are different families of locality-sensitive hash functions for different distance or similarity measures, with some of the most popular being \emph{simhash}~\cite{Charikar2002}, \emph{MinHash}~\cite{BCFM2000} and \emph{sdhash}~\cite{Roussev2010}.
For example, the MinHash of a set is the minimum element according to a uniformly random permutation.
The probability that two sets $A$ and $B$ hash to the same value is precisely their Jaccard similarity $(|A \cap B|)/(|A \cup B|)$. 
%As the collision probability can be appropriately upper- and lower-bounded for chosen similarity thresholds, MinHash is locality-sensitive w.r.t. Jaccard similarity.

\section{Delayed Deterministic Finite Automata}\label{sec:ddfa_model}
\begin{figure}[t]
    \centering
    \includegraphics[width=\textwidth]{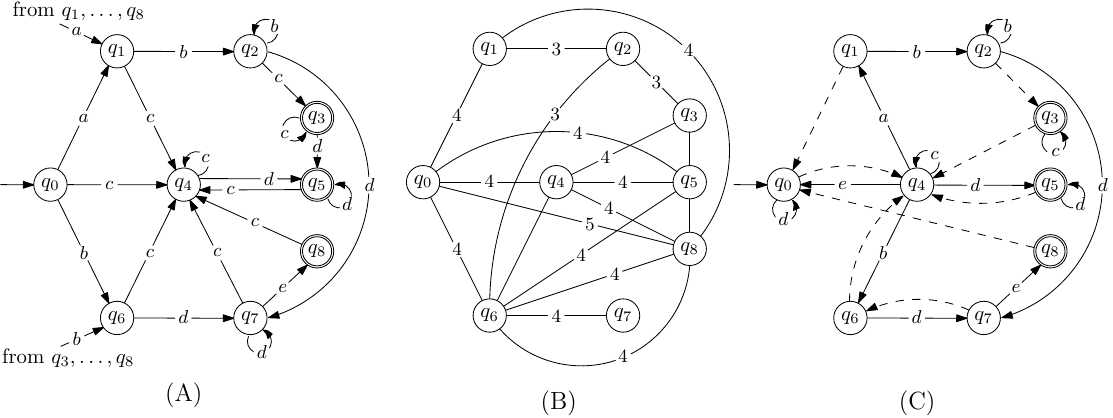}
    \caption{Example from~\cite{KDYP+2006}. 
        (A) DFA $D$ for regular expression \texttt{.*((ab+c+)|(cd+)|(bd+e))}. Edges to $q_0$ are omitted. 
        (B) Space reduction graph for $D$ with edges annotated with similarity. Edges with similarity less than $4$ omitted, except those connecting $q_2$ to avoid disconnecting the graph. 
        (C) \ddfa{} equivalent to $D$. All transitions are shown, default transitions are dashed.}
    \label{fig:example}
\end{figure}

A \emph{delayed deterministic finite automaton} (\ddfa{})~\cite{KDYP+2006} is a deterministic finite automaton that is augmented with unlabeled \emph{default transitions}.
Formally a \ddfa{} is a 6-tuple $D^2 = (Q, \Sigma, \delta, q_0, A, F)$.
As for DFAs, $Q$ is the set of states, $\Sigma$ is the alphabet, $\delta$ is the transition functions, $q_{0} \in Q$ is the initial state, and $A \subseteq Q$ is the set of accepting states. The final component $F: Q \rightarrow Q$ is the \emph{default transition function}. Viewed as a graph, default transitions are $\epsilon$-labeled directed edges, where $\epsilon$ is the empty string, and each state has at most one outgoing default transition. See Figure~\ref{fig:example}~(C) for an example.

To transition from a state $u$ according to a character $c$, we follow a $c$-labeled transition if it exists or otherwise follow the default transition:
\[
  \delta(u, c) =
    \begin{cases}
        v &\text{ if } (u, v)_{c} \text{ is a edge.} \\
        \delta(F(u), c) &\text{ otherwise }
    \end{cases}
\]
Note that for $\delta$ to be well-defined, it must always be possible to reach a state from $u$ with a $c$-labeled transition.
This implies that any cycle of default transitions must have an outgoing $c$-labeled transition for any character $c$.
To transition from a state $u$ according a string $S = c_{1} \ldots c_{m}$ we recursively transition according to each character:
\[ \delta(u, c_{1} c_{2} \ldots c_{m}) = \delta(\delta(u, c_{1}), c_{2} \ldots c_{m}). \]
Given a character $c$ and a path $p = u_{1}, \ldots, u_{k}$ we say $p$ matches $c$ if $\delta(u_{1}, c) = u_{k}$ and all but the last transition is default, i.e., $F(u_{i}) = u_{i+1}$ for $1 \leq i < k$.
Note that the concatenation of the labels of $p$ equals $c$.
%Given a character $c$ and a path $p = u_{1}, \ldots, u_{k}$ we say $p$ matches $c$ if $\delta(u_{1}, c) = u_{k}$ and the concatenation of the labels of $p$ equals $c$, which implies that all but the last transition is default, that is, $F(u_{i}) = u_{i+1}$ for $1 \leq i < k$.
Given a string $S = c_{1} \ldots c_{m}$ and a path $p$ we say $p$ matches $S$ if $p$ is the concatenation of paths matching the individual characters $c_{1}, \ldots, c_{m}$. Note that the concatenation of the labels of $p$ is $S$. We define acceptance as before. Two \ddfa{}s are equivalent if they have the same language, and two transitions are equivalent if they have the same destination and label.

Given a DFA, we can compress it by replacing sets of equivalent transitions with single default transitions to obtain an equivalent \ddfa{} with fewer total transitions. We define the \emph{similarity} of two states $u$ and $v$, denoted $\simi(u, v)$, to be their number of equivalent transitions, that is, $\simi(u, v) = |\{ c \in \Sigma \mid \delta(u,c) = \delta(v,c) \}|$.
See Figure~\ref{fig:example}~(B) for an illustration.
Inserting a default transition $(u, v)$ and removing the equivalent transitions from $u$ does not affect the language but saves $\simi(u, v) - 1$ transitions.
Each transition we can remove without affecting the language we say is \emph{redundant}.

Following a default transition does not consume an input character, which introduces a \emph{delay} when matching. We define the \emph{longest delay} of $D^{2}$ to be maximum number of default transitions in any path matching a single character, i.e., the longest delay is the maximum number of default transitions in $D^{2}$ to match any single character. 
%That is, if the longest delay is $d$ we must follow at most $d$ default transitions to match any single character. 
Given a string $S$, we define the \emph{matching delay} of $S$ in $D^{2}$ to be the number of default transitions in the path starting in $q_{0}$ and matching $S$.

\section{Compressing DFAs with Default Transitions}\label{sec:original_ddfa_construction}
We now review the algorithm of Kumar~et~al.~\cite{KDYP+2006} that compress a DFA $D$ into an equivalent \ddfa{}.  
Let $D^{2}$ be an initially empty \ddfa{} with the same set of states as $D$. We proceed as follows.
\begin{description}
    \item [Step 1: Space Reduction Graph]
    Construct the complete, undirected  graph over the states of $D$, and to each edge $(u, v)$ assign weight $\simi(u, v)$.
    This is the \emph{space reduction graph} (SRG).
    See Figure~\ref{fig:example}~(B) for an example.

    \item [Step 2: Maximum Spanning Tree]
    Build a maximum spanning tree over the SRG.
    Root the spanning tree in a central node, i.e., a node of minimal radius, and direct all edges towards the root to obtain a directed spanning tree.

    \item [Step 3: Transitions]
    For each edge $(u, v)$ in the tree insert the default transition $(u, v)$ into $D^{2}$.
    Copy every labeled transition from $D$ into $D^{2}$ that is not redundant in $D^{2}$.
\end{description}
Since the SRG weighs the edges by similarity, computing the maximum spanning tree maximizes the overall compression. 

Step~1 uses $O(n^{2} |\Sigma|)$ to compare all states and construct the SRG. Step~2 constructs the maximum spanning tree with Kruskal's algorithm~\cite{Kruskal1956} in $O(n^{2} \log n)$ time, and finally step 3 takes $O(n |\Sigma|)$ time. In total, the running time is $O(n^{2} \log n + n^{2}|\Sigma|))$\footnote{The running time is not explicitly stated in the paper, but follows from the description.}. 

\section{Fast Compression}\label{sec:fast_ddfa_construction}
We now show how to speed up the DFA compression algorithm using locality-sensitive hashing to sparsify the SRG construction in step 1. 

Let $D$ be the input DFA and let $r$ and $k$ be two constant, positive integer parameters. We initialize an undirected graph $G = (Q, E)$ where $Q$ is the set of states in $D$ and $E = \{ (q_{0}, u)\ |\ u \neq q_{0}\}$, i.e., we have a edge between the initial state and all other states. We then add edges to the graph in $r$ rounds, where each round proceeds as follows: 

First, pick $k$ unique random characters $c_1, \ldots, c_k \in \Sigma$. Then, for each state $v \in Q$ we construct the sequence of $k$ states $V = \delta(v, c_{1}), \ldots, \delta(v, c_{k})$. We hash $V$ into a single hash value $h(v)$ using a standard hashing scheme of Black~et~al.~\cite{BHKK+1999}. We insert $v$ into a table with key $h(v)$. For each unique hash value $h_i$, consider the set of states $C_i$ that hash to $h_i$. For each state $u \in C_{i}$, we pick another state $v \in C_i$ uniformly at random and insert $(u, v)$ into $E$ if it does not already exist.

After $r$ rounds, the algorithm terminates, and we assign weights to each edge of $G$ equal to the similarity of the endpoint states. The resulting graph $G$ is the \emph{sparse space reduction graph} (SSRG). 

The above scheme is inspired by the locality-sensitive scheme by Har-Peled~et~al.~\cite{PIM2012} for bitstrings w.r.t. Hamming distance. Their solution samples positions from the input bitstrings. In our solution, the sampled positions correspond to the sampled characters from $\Sigma$. 

Hashing a state takes $O(k)$ time, and sampling an edge takes constant time. Thus, a round takes $O(kn)$ time and we use $O(rkn)$ time for all rounds. Each round inserts at most $n$ edges and hence the SSRG $G$ has $O(rn)$ edges at the end. We compute the similarity between two states in $O(|\Sigma|)$ time, and hence the final algorithm uses $O(rkn + rn |\Sigma|) = O(n |\Sigma|)$ time. 

We plug in the modified Step 1 in the algorithm from Section~\ref{sec:original_ddfa_construction}. Step 1 takes $O(n |\Sigma|)$ time and since $G$ has $O(n)$ edges, Step 2 takes $O(n \log n)$ time. Step 3 takes $O(n |\Sigma|)$ time as before. In total, we use $O(n \log n + n|\Sigma|)$ time.

\section{Compression with Bounded Longest Delay}\label{sec:worst_case_overhead}
We now consider the bounded longest delay variant of \ddfa{}, that is, given a DFA $D$ and an integer parameter $L$, construct a \ddfa{} $D^2$ equivalent to $D$, such that the longest delay of $D^2$ is at most $L$. Recall that the longest delay of a \ddfa{} is the length of the longest path of default transitions in $D^2$. 

\subsection{Bounded Longest Delay by Constructing Small Trees}\label{sec:limit_refined}
We first review the algorithm by Kumar~et~al.~\cite{KDYP+2006}. The algorithm is based on a simple modification to the maximum spanning tree construction in Step 2 of the algorithm from Section~\ref{sec:original_ddfa_construction}. 

Let $L$ be a parameter. We modify Step 2 by constructing a maximum spanning forest with the constraint that each tree in the forest has a diameter of at most $\Delta = 2L$. To do so, we run Kruskal's algorithm but simply ignore any edges that would cause a tree diameter to exceed $\Delta$. Also, among the edges with maximum similarity, we select one that causes a minimum increase to any tree diameter. After constructing the forest, we root each tree in a central node and direct edges toward each root. 

Since each tree has diameter at most $\Delta$ and is rooted in a central node, the final \ddfa{} has a longest delay of at most $\lceil \Delta / 2 \rceil = L$. To implement the modified Step 2, Kumar~et~al.~\cite{KDYP+2006} maintains the radius of the tree for each node during the maximum spanning forest construction. When we add an edge, we need to merge two trees and potentially update the radius for each node in the resulting tree. Hence, we may need to update $\Omega(n^2)$ radii in total during the maximum spanning forest construction. 

We note that new Step 2 uses $\Omega(n^2)$ time, whether or not we run it on a sparse or a dense space reduction graph. Hence, we cannot directly apply our sparsification technique from Section~\ref{sec:fast_ddfa_construction} to obtain an efficient algorithm. We present a new algorithm in the next section that efficiently combines with sparsification.

\subsection{Fast Compression with Bounded Longest Delay}\label{sec:limit_cut}
We now present a new algorithm that uses sparsification to efficiently construct \ddfa{}s with bounded longest delay. The idea is to construct a single large maximum spanning tree and then cut edges until each tree in the resulting forest has small diameter.

Given a DFA $D$ and an integer parameter $L$, we proceed as follows. 
\begin{description}
    \item [Step 1: Construct SSRG]
        Construct a sparse space reduction graph $G$ for $D$ as in Section~\ref{sec:fast_ddfa_construction}.
    \item [Step 2: Construct MST]
        Construct a maximum spanning tree $T_0$ over $G$ using Kruskal's algorithm.
        Pick a central node~$v_{0}$ in $T_0$ and then discard $T_0$.
        Construct a new maximum spanning tree $T$ using Prim's algorithm~\cite{Prim1957} with~$v_{0}$ as the initial node. When we queue a new edge $(u, v)$, where~$u$ is the node already in~$T$, assign weight $w'_{u, v} = \simi(u, v) - 2^{d_{v}}$ where~$d_{v}$ is the distance from~$v_{0}$ to~$v$ in~$T$.
    \item [Step 3: Cut Edges]
        Cut a minimum number of edges in $T$ to obtain a forest with each tree diameter at most~$\Delta = 2L$. We do so using the algorithm of Farley~et~al.~\cite{FHP1981} that cuts the necessary edges in a bottom-up traversal of $T$. Then, direct the edges of each tree towards the root.
    \item [Step 4: Construct Transitions]
        As in Step~3 in Section~\ref{sec:original_ddfa_construction}, create default transitions along edges in the trees and then copy in every labeled transition from $D$ that is not redundant in $D^2$.
\end{description}
After cutting, each tree has a diameter of at most $\Delta = 2L$. Since we direct edges toward each root, the longest delay is at most $\lceil \Delta / 2 \rceil = L$. Step~1 and 4 uses $O(n|\Sigma|)$ time as before. Step 2 uses $O(n \log n)$ time, and Step 3 uses $O(n)$ time. In total, we use $O(n \log n + n |\Sigma|)$ time.

%As before, we construct the SRG in $O(rn(k + |\Sigma|))$ time.
%Prim's algorithm has the same time complexity as Kruskal's, so we construct both $T_0$ and $T$ in $O(m \log m) = O(rn + \log rn)$ time, where $m = O(rn)$ is the number of edges in the SRG.
%We use the algorithm of Farley~et~al.~\cite{FHP1981} to the cut edges by a bottom up traversal of $T$ in $O(n)$ time.
%Thus the total construction time is $O(rn(k + |\Sigma| + \log rn)) = O(n (|\Sigma| + \log n))$.

Each edge $(u, v)$ we cut results in $\simi(u, v) - 1$ more labeled transitions in $D^2$, as that default transition is then not constructed. Intuitively, the lower the diameter of $T$, the fewer edges we cut to get each tree below the bound. Therefore, we use an edge weight that trades similarity for lower diameter, as we observed that the fewer cuts outweighed the lost similarity. 

We found the simple heuristic of the modified weight $w'$ performed well in practice. A similar idea was used by Kumar~et~al~\cite{KDYP+2006} in their solution. 

Because $T$ is not a maximum spanning tree w.r.t. similarity, the choice of initial node $v_{0}$ affects the total similarity of the final tree. We found that picking $v_{0}$ to be a central node in a MST w.r.t. similarity ($T_0$) yielded the best compression in practice.

Note that we cut the minimum \emph{number} of edges to uphold the diameter constraint.
Alternatively, we could cut edges of minimum \emph{total similarity}, which might result in better compression. However, our approach is simple and fast in practice, and because each edge in the SRG has near-maximum weight, the difference in compression is negligible.
%We note that cutting edges such that the \emph{total similarity} is minimized would be optimal, but we did not find any sufficiently fast algorithm that solves this problem.
%Instead we cut a minimum \emph{number} of edges, which works well in practice as most edges in the SRG have near-maximum weight.

\section{Compression with Bounded Matching Delay}\label{sec:adfa}
We now consider the bounded matching delay variant of \ddfa{}, that is, given a DFA $D$, construct a \ddfa{} $D^2$ equivalent to $D$, such that the matching delay of $D^2$ is at most $|S|$ on any input string $S$. 

\subsection{Bounded Matching Delay by the A-DFA Algorithm}\label{sec:adfa_construction}
We first review \adfa{} algorithm by Becchi and Crowley~\cite{BC2007b, BC2013}. Let $D$ be an input DFA. For a state $v \in Q$, define the \emph{depth} of $v$, denoted $d(v)$, to be the length of the shortest path from the initial state $q_{0}$ to $v$. The key idea is only to add default transitions from state $v$ to state $u$ if $d(u) < d(v)$. This implies that the matching delay is at most $|S|$ on any input string $S$ (see, e.g., Aho and Corasick~\cite{AC1975}). 

Initialize a \ddfa{} $D^{2}$ with no default transitions. We proceed as follows.
\begin{description}
    \item [Step 1: Calculate Depth]
        Calculate the depth $d(v)$ of each state $v \in Q$ by a breadth-first traversal of~$D$.
    \item [Step 2: Construct Default Transitions]
        For each state $u \in Q$ add default transition $(u, v)$ to $D^2$, where $v$ is the state such that $\simi(u, v)$ is maximum and $d(v) < d(u)$.
    \item [Step 3: Construct Labeled Transitions]
        Copy in every labeled transition from $D$ that is not redundant in $D^2$.
\end{description}
Step~1 and~3 uses $O(n |\Sigma|)$ time to traverse $D$. Step~2 uses $O(n^2 |\Sigma|)$ to compute the similarity of each pair of states. In total, we use $O(n^2 |\Sigma|)$ time.

%Note that the algorithm does not construct the complete SRG. However, selecting a maximum-similarity edge in Step~2 is equivalent to selecting a maximum-weight incident edge in the complete SRG (that also leads to lower-depth state), and asymptotically takes as much time as constructing those edge beforehand. Therefore this algorithm can be thought of as \emph{implicitly} using the SRG.

\subsection{Fast Compression with Bounded Matching Delay}\label{sec:adfa_construction_sparse}
We now speed up the \adfa{} algorithm using sparsification. Let $D$ be the input DFA and let $r$ and $k$ be two constant, positive integer parameters. We initialize a \ddfa{} $D^2$ with no default transitions, i.e., we set $F(u) = u$ for each $u \in Q$. The algorithm runs in $r$ rounds, where each round proceeds as follows.

First, pick $k$ unique random characters $c_1, \ldots, c_k \in \Sigma$. Then, for each state $v \in Q$ we construct the sequence $V = \delta(v, c_{1}), \ldots, \delta(v, c_{k})$, and hash $V$ into a single hash value $h(v)$. We insert $v$ into a table with key $h(v)$.
For each unique hash value~$h_i$, we consider the set of states $C_i$ that hash to that value.
For each state $u \in C_{i}$ we pick another state $v \in C_i$ uniformly at random.
If $v$ has lower depth and the default transition $(u, v)$ compresses better than the current default transition of $u$, i.e., $d(v) < d(u)$ and $\simi(u, v) > \simi(u, F(u))$, we update the default transition of~$u$ to point to~$v$ in $D^2$, otherwise, we continue. After~$r$ rounds, the algorithm terminates and returns the resulting \ddfa{} $D^2$.

Hashing a state and computing the similarity of the potential new default transition uses $O(k |\Sigma|)$ time. Hence, the full algorithm uses $O(rkn + rn|\Sigma|) = O(n |\Sigma|)$ time. 

\section{Experimental Evaluation}\label{sec:experiments}
We implemented our methods described in the previous section and measured their performance on regular expressions extracted from widely used intrusion detection systems.  The implementation is available at \texttt{https://github.com/MaxRishoj/fcomp-dfa}.

\subsection{Datasets}
%We extracted updated versions of the datasets used in prior work from publicly available regular expression sets in real-world intrusion detection systems.
We extracted our datasets from regular expressions used in the popular Snort~\cite{Roesch1999}, Zeek(formerly Bro)~\cite{Paxson1999}, and Suricata~\cite{suricata2010} intrusion detection systems (see current homepages for these systems at \url{www.snort.org}, \url{zeek.org}, and \url{suricata.io}). The Snort and Zeek datasets are extracted from current versions of the datasets used in most of the previous work.

For each dataset, we extracted prefixes of the rules to generate DFAs of different sizes to explore the scalability of our algorithms on DFAs that have between roughly $1$k to $1$M states. See Table~\ref{tab:rule_characteristics} in the appendix for details. As in previous work, we have filtered out some rules that used advanced features. All regular expressions use an ASCII alphabet size of $256$.

\subsection{Algorithms Tested}
We evaluate the following algorithms.  
\begin{description}[font=\normalfont,style=standard]
 \item[\algoriginalfull{}.] The algorithm of~\cite{KDYP+2006}, described in Section~\ref{sec:original_ddfa_construction}. 
    \item[\algrefinedfull{}.] The algorithm of~\cite{KDYP+2006} for bounded longest delay, described in Section~\ref{sec:limit_refined} with parameter $L = 2$. 
    \item[\algcutfull{}.] The algorithm presented in Section~\ref{sec:limit_cut} for bounded longest delay using the SRG instead of the SSRG.
    \item[\algadfafull{}.] The algorithm of~\cite{BC2007b, BC2013} for bounded matching delay, described in Section~\ref{sec:adfa}.  
     \item[\algoriginalsparse{}.] The algorithm presented in Section~\ref{sec:fast_ddfa_construction}. 
    \item[\algrefinedsparse{}.] As \algrefinedfull{} using the SSRG instead of the SRG.
    \item[\algcutsparse{}.] The algorithm presented in Section~\ref{sec:limit_cut} for bounded longest delay. 
    \item[\algadfasparse{}.] The algorithm presented in Section~\ref{sec:adfa_construction_sparse} for bounded matching delay.
\end{description}

We use the locality-sensitive hashing scheme from Section~\ref{sec:fast_ddfa_construction} without replacement and parameters $k = 8$ and $r = 512$. We evaluated several locality-sensitive hashing schemes, including the one in Section~\ref{sec:fast_ddfa_construction} with replacement and minhash~\cite{Broder1997} over the set of outgoing transitions with one or $k$ random permutations of the universe. We also evaluated several combinations of $r$ and $k$ and found that $k=8$ gave the best compression size. Increasing $r$ results in better compression size but increases the compression time linearly. Our chosen variant achieved the best combination of compression size and compression time. 

In our longest delay variant, we report results for parameter $L = 2$. We have experimented with other values of $L$ but did not observe significant differences in the relative performances of the algorithm for this variant. We note that the bounded matching delay variant also appears in a more general version, where we can tune the overhead of the default transitions (see the journal version of the result~\cite{BC2013}). The version tested here is the simplest version and leads to the best compression size.  

\subsection{Setup}
Experiments were run on a machine with an Intel Xeon Gold 6226R 2.9GHz processor and $128$GB of memory. The operating system was Scientific Linux 7.9 kernel version 3.10.0-1160.80.1.el7.x86\_64.
Source code was compiled with \texttt{g++} version 9.4 with options \texttt{-Wall -O4}.
The input to each algorithm is a DFA constructed from a set of regular expressions.
We measured the time for constructing an equivalent \ddfa{} for the input DFA, using the \texttt{clock} function of the C standard library. 

\subsection{Results}
\begin{figure}
    \centering
    \includegraphics[width=\linewidth]{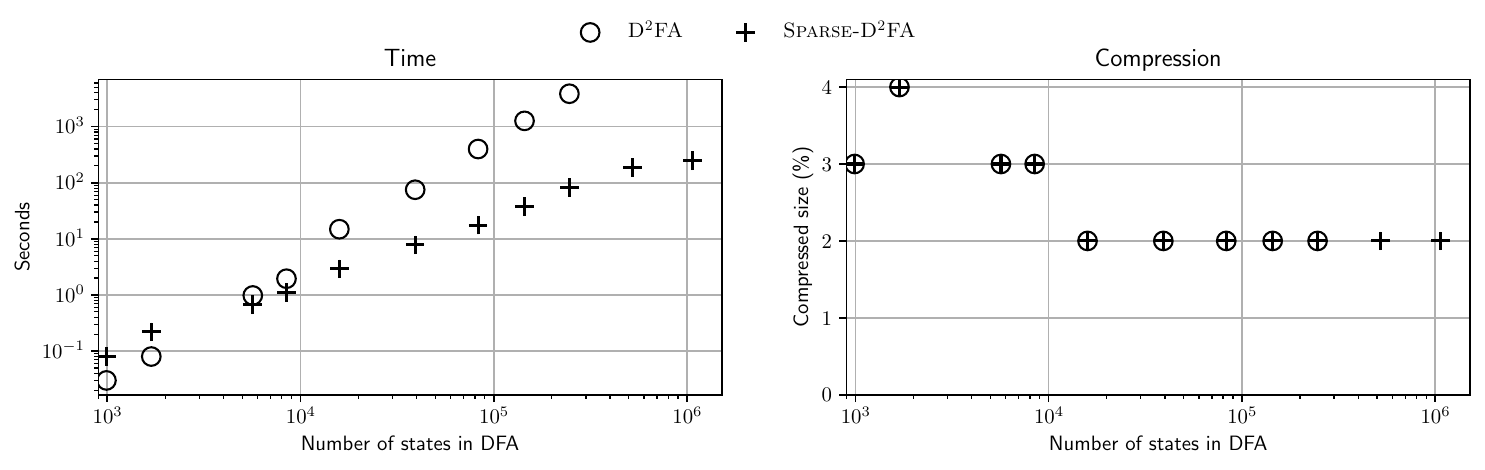}
    \\ \vspace{1\baselineskip}
    \includegraphics[width=\linewidth]{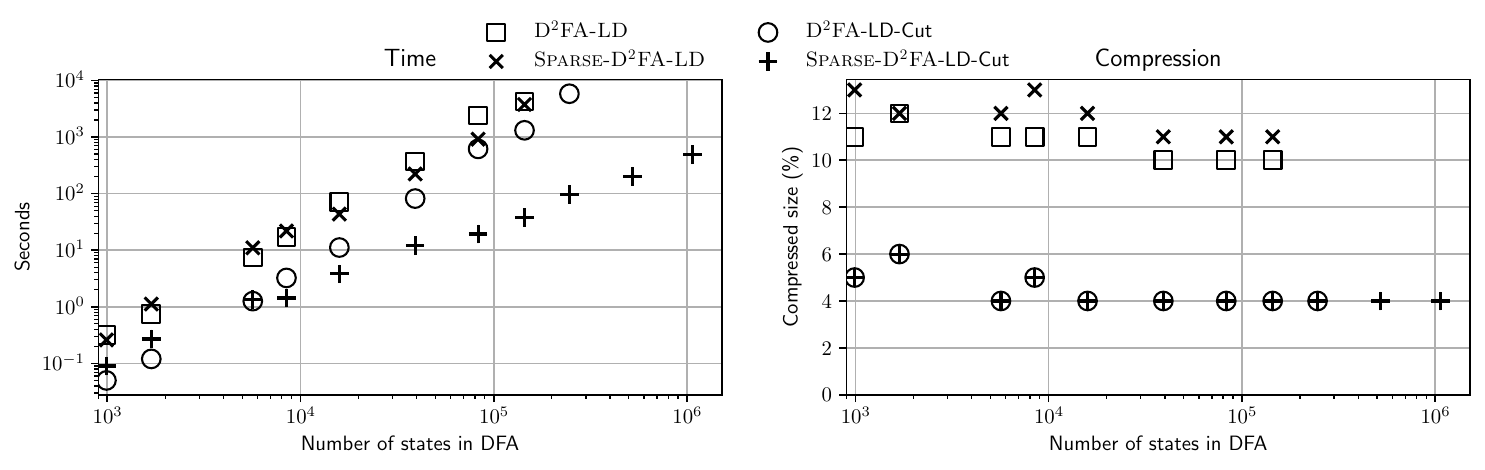}
    \\ \vspace{1\baselineskip}
    \includegraphics[width=\linewidth]{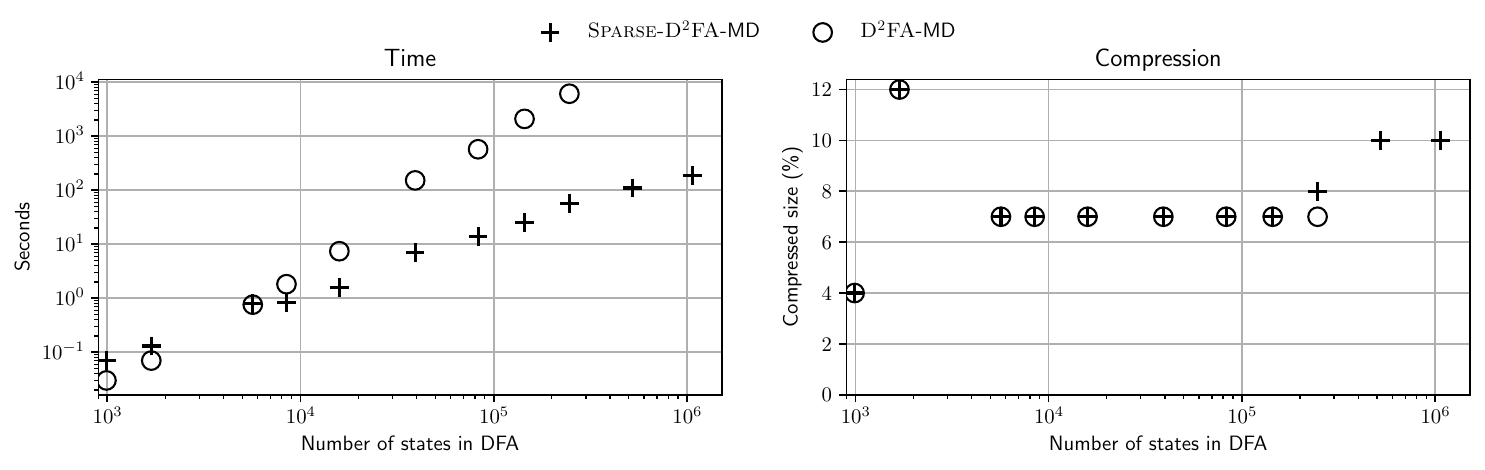}

    \caption{Results for the Snort dataset on the algorithms for general compression (top), bounded longest delay (middle), and bounded longest matching delay (bottom). On the left, we show compression time in seconds vs. the number of states in the input DFA. On the right, we show the number of transitions in the \ddfa{} as a percent of the number of transitions in the input DFA.} 
    \label{fig:snort}
\end{figure}

\begin{figure}
    \centering

   \includegraphics[width=\linewidth]{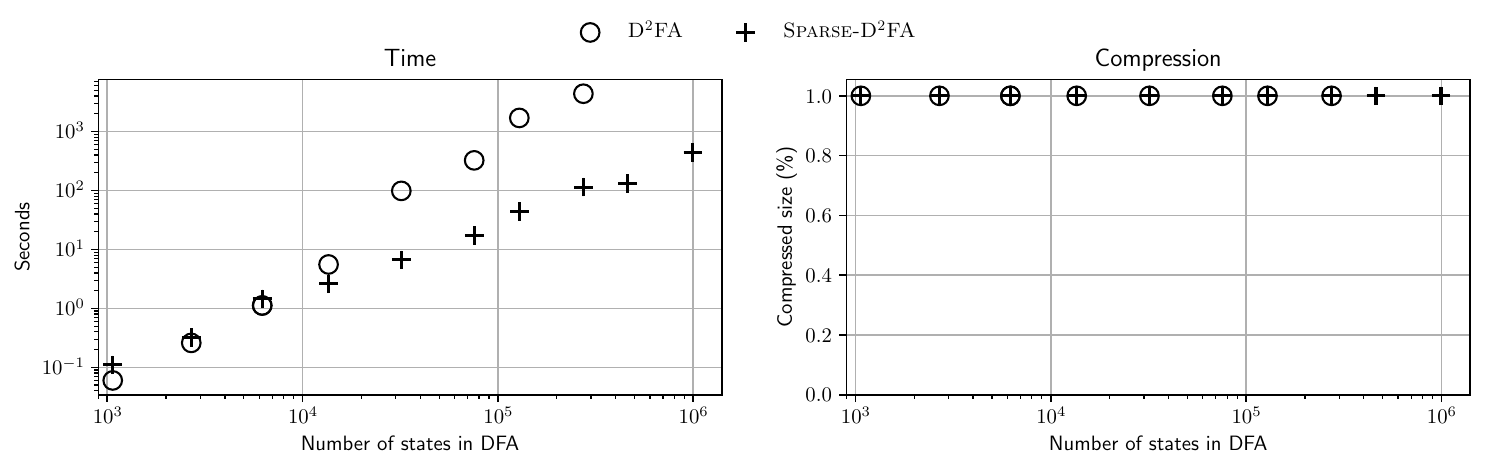}
    \\ \vspace{1\baselineskip}
    \includegraphics[width=\linewidth]{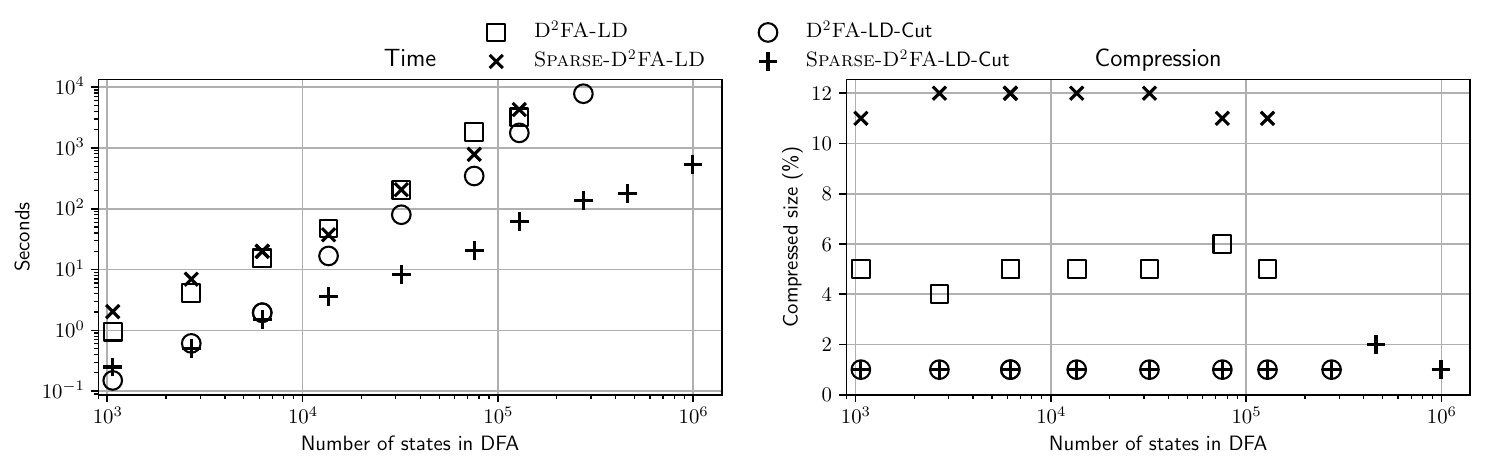}
    \\ \vspace{1\baselineskip}
    \includegraphics[width=\linewidth]{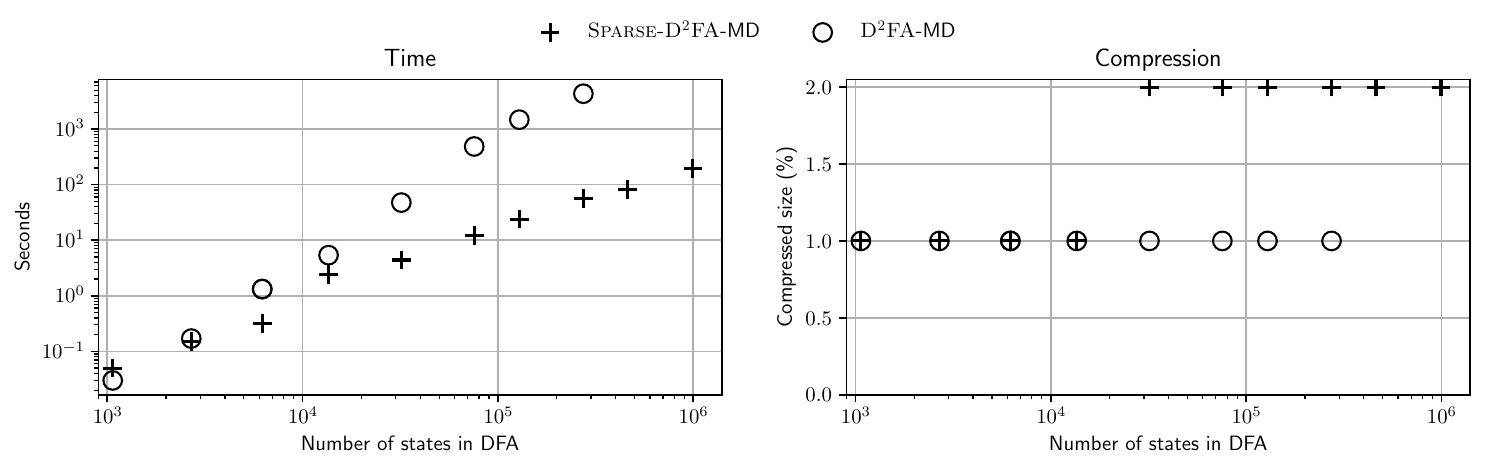}
    
   \caption{Results for the Suricata dataset on the algorithms for general compression (top), bounded longest delay (middle), and bounded longest matching delay (bottom). On the left, we show compression time in seconds vs. the number of states in the input DFA. On the right, we show the number of transitions in the \ddfa{} as a percent of the number of transitions in the input DFA.} 
    \label{fig:suricata}
\end{figure}

\begin{figure}
    \centering

    \includegraphics[width=\linewidth]{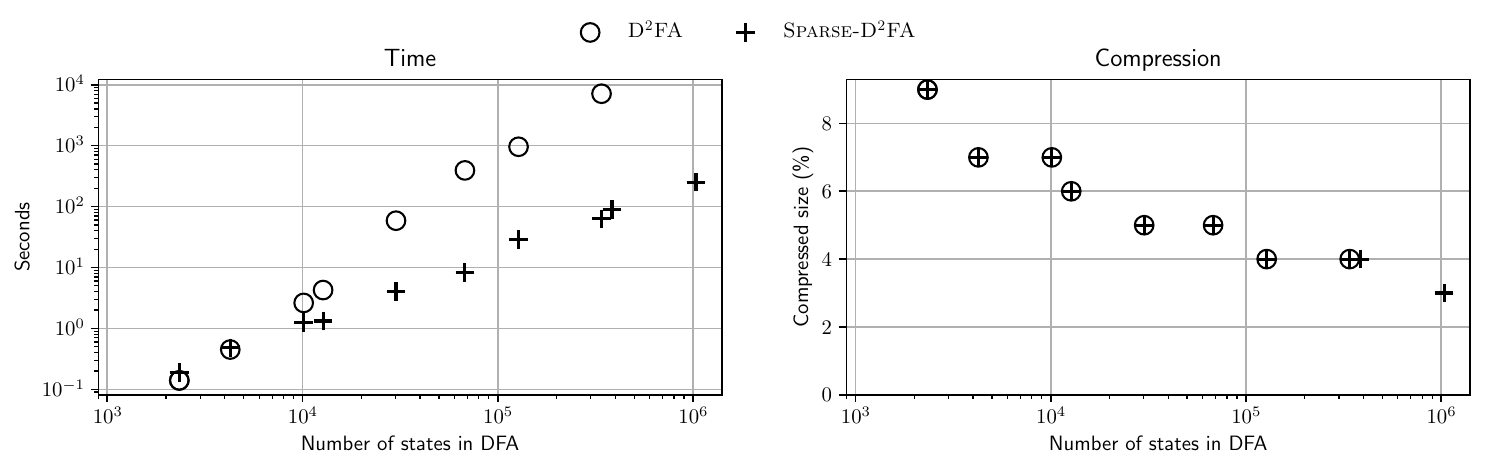}
    \\ \vspace{1\baselineskip}
    \includegraphics[width=\linewidth]{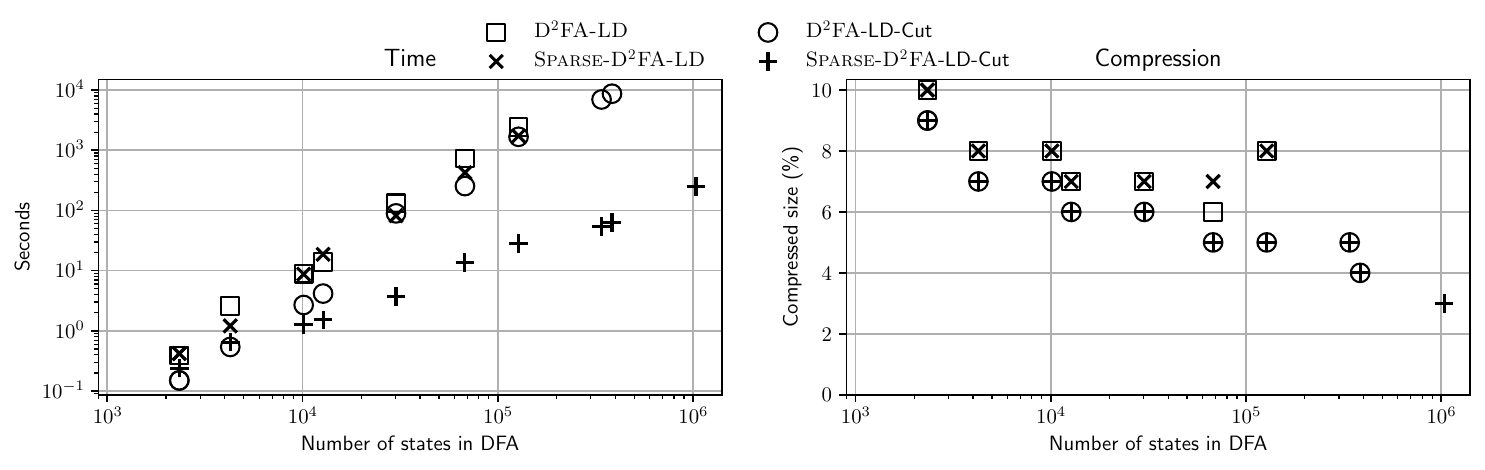}
    \\ \vspace{1\baselineskip}
    \includegraphics[width=\linewidth]{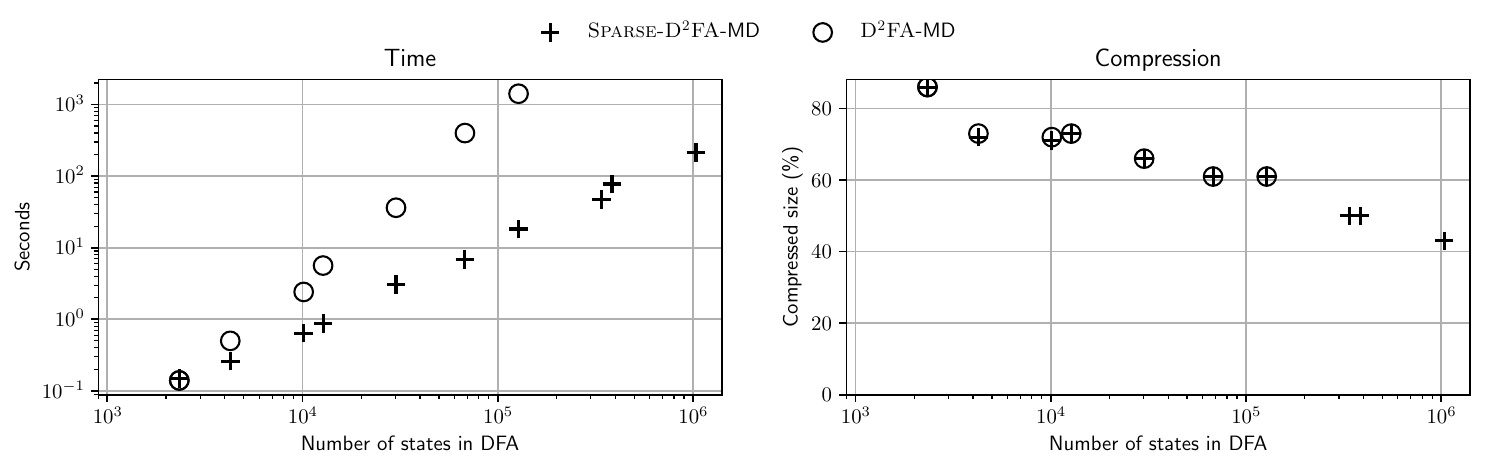}
    
   \caption{Results for the Zeek dataset on the algorithms for general compression (top), bounded longest delay (middle), and bounded longest matching delay (bottom). On the left, we show compression time in seconds vs. the number of states in the input DFA. On the right, we show the number of transitions in the \ddfa{} as a percent of the number of transitions in the input DFA.} 
    \label{fig:zeek}
\end{figure}

We compare the algorithms across the datasets and measure compression time and compression size (number of states in \ddfa{} as a percent of the number of states in the input DFA) for each of the variants (general compression, longest delay, and bounded matching delay). The relative performance of our algorithms is similar across the datasets, so we focus on the results for the Snort dataset shown in Figure~\ref{fig:snort}. The corresponding results for the Suricata and Zeek datasets are in  Figures~\ref{fig:suricata} and~\ref{fig:zeek}. We point out whenever there are significant differences between observed results across datasets. We note that the absolute compression size varies significantly across the datasets and variants. Most instances are highly compressible to around 10\% of the original DFA (many even in the low single-digit percentages). At the same time, a single one (bounded matching delay on the Zeek dataset) compresses to around $50$ percent on the largest DFAs. 

\paragraph{General Compression}
We compare the \algoriginalfull{} and \algoriginalsparse{} general compression algorithms. We observe that \algoriginalsparse{} compresses up to an order of magnitude faster than \algoriginalfull{} with either no or little loss of compression size. For DFAs with around $1$k states, the compression time is comparable and increases gradually to roughly an order of magnitude for the largest DFAs.

\paragraph{Compression with Bounded Longest Delay}
We compare the \algrefinedfull{}, \algrefinedsparse{}, \algcutfull{} and \algcutsparse{} compression algorithms with bounded longest delay. We observe that \algrefinedfull{} and  \algrefinedsparse{} achieve similar compression times as expected. The compression size achieved by \algrefinedsparse{} is around 10-15\% worse in the Snort dataset, roughly comparable in the Zeek dataset, and around 100\% worse in the Suricata dataset. We believe that the bounded diameter approach in Kumar et al.~\cite{KDYP+2006} is highly sensitive to the greedy choice of edges at each step, leading to the observed differences in the compression size across the different datasets.    

More importantly, we observe that \algcutsparse{} compresses up to an order of magnitude faster than all other algorithms. \algcutsparse{} also achieves a substantially better compression size than  \algrefinedfull{} and \algrefinedsparse{} (except for the Zeek dataset, where the compression size is comparable). Compared to the dense version \algcutfull{}, \algcutsparse{} has either no or little loss of compression size.

\paragraph{Compression with Bounded Matching Delay}
We compare the compression algorithms \algadfafull{} and \algadfasparse{} with bounded matching delay. We observe that \algadfasparse{} compresses up to an order of magnitude faster than \algadfafull{}. The compression size varies depending on the dataset. For the Snort dataset, \algadfasparse{} achieves 10-15\% worse compression size for DFAs with more than $10$k states, for the Zeek dataset, \algadfasparse{} is comparable, and for the Suricata dataset, \algadfasparse{} achieves 100\% worse compression size for DFAs with more than $10$k states. We note that the absolute compression size varies significantly, which may explain this difference in relative compression size. 

\section{Conclusion}\label{sec:conclusion}
We presented a simple, general framework for constructing \ddfa{} based on locality-sensitive hashing that constructs an approximation of the optimal \ddfa{} in near-linear time and applied the approach to the \ddfa{} compression algorithm and two important variants. On DFAs from widely used modern intrusion detection systems, we achieved compression times of up to an order of magnitude faster than existing solutions with either no or little loss of compression size.

An interesting open problem is to explore if our new framework can be combined with other  DFA compression techniques, such as the ones mentioned in Section~\ref{sec:related_work}. Substantial work has been done on compression DFAs, and we believe it is likely that combinations with our technique will lead to even faster solutions. 

\section{Acknowledgments}
This paper is inspired by earlier work in an MSc. thesis~\cite{BM2020} supervised by two of the authors. We thank the anonymous reviewers of earlier drafts of this article for many valuable comments that improved the quality of the work.

\bibliographystyle{plainurl}
\bibliography{references}

\begin{thebibliography}{10}

\bibitem{suricata2010}
www.suricata.io.

\bibitem{AC1975}
Alfred~V. Aho and Margaret~J. Corasick.
\newblock Efficient string matching: an aid to bibliographic search.
\newblock {\em Commun. ACM}, 18(6):333--340, 1975.
\newblock \href {https://doi.org/http://doi.acm.org/10.1145/360825.360855}
  {\path{doi:http://doi.acm.org/10.1145/360825.360855}}.

\bibitem{AFSK+2012}
Rafael Antonello, Stenio F.~L. Fernandes, Djamel Sadok, Judith Kelner, and
  G{\'{e}}za Szab{\'{o}}.
\newblock Deterministic finite automaton for scalable traffic identification:
  The power of compressing by range.
\newblock In {\em {NOMS} 2012}, pages 155--162, 2012.
\newblock \href {https://doi.org/10.1109/NOMS.2012.6211894}
  {\path{doi:10.1109/NOMS.2012.6211894}}.

\bibitem{AFSK+2015}
Rafael Antonello, Stenio F.~L. Fernandes, Djamel Fawzi~Hadj Sadok, Judith
  Kelner, and G{\'{e}}za Szab{\'{o}}.
\newblock Design and optimizations for efficient regular expression matching in
  {DPI} systems.
\newblock {\em Comput. Commun.}, 61:103--120, 2015.
\newblock \href {https://doi.org/10.1016/j.comcom.2014.12.011}
  {\path{doi:10.1016/j.comcom.2014.12.011}}.

\bibitem{BC2007a}
Michela Becchi and Srihari Cadambi.
\newblock Memory-efficient regular expression search using state merging.
\newblock In {\em Proc. 26th {INFOCOM}}, pages 1064--1072, 2007.
\newblock \href {https://doi.org/10.1109/INFCOM.2007.128}
  {\path{doi:10.1109/INFCOM.2007.128}}.

\bibitem{BC2007}
Michela Becchi and Patrick Crowley.
\newblock A hybrid finite automaton for practical deep packet inspection.
\newblock In {\em Proc. 3rd {CoNEXT} conference}, pages 1--12, 2007.

\bibitem{BC2007b}
Michela Becchi and Patrick Crowley.
\newblock An improved algorithm to accelerate regular expression evaluation.
\newblock In {\em Proc. {ANCS} 2007}, pages 145--154, 2007.
\newblock \href {https://doi.org/10.1145/1323548.1323573}
  {\path{doi:10.1145/1323548.1323573}}.

\bibitem{BC2008}
Michela Becchi and Patrick Crowley.
\newblock Efficient regular expression evaluation: theory to practice.
\newblock In {\em Proc. {ANCS} 2008}, pages 50--59, 2008.
\newblock \href {https://doi.org/10.1145/1477942.1477950}
  {\path{doi:10.1145/1477942.1477950}}.

\bibitem{BC2013}
Michela Becchi and Patrick Crowley.
\newblock {A-DFA:} {A} time- and space-efficient {DFA} compression algorithm
  for fast regular expression evaluation.
\newblock {\em {ACM} Trans. Archit. Code Optim.}, 10(1):4:1--4:26, 2013.
\newblock \href {https://doi.org/10.1145/2445572.2445576}
  {\path{doi:10.1145/2445572.2445576}}.

\bibitem{BGP2025}
Philip Bille, Inge~Li G{\o}rtz, and Max~Rish{\o}j Pedersen.
\newblock Fast practical compression of deterministic finite automata.
\newblock In {\em Proc. 50th SOFSEM}, 2025.

\bibitem{BGPT2023}
Philip Bille, Inge~Li G{\o}rtz, Simon~J. Puglisi, and Simon~R. Tarnow.
\newblock Hierarchical relative lempel-ziv compression.
\newblock In {\em Proc. 21st {SEA}}, 2023.

\bibitem{BHKK+1999}
John Black, Shai Halevi, Hugo Krawczyk, Ted Krovetz, and Phillip Rogaway.
\newblock {UMAC:} fast and secure message authentication.
\newblock In {\em Proc. 19th {CRYPTO}}, volume 1666, pages 216--233, 1999.
\newblock \href {https://doi.org/10.1007/3-540-48405-1\_14}
  {\path{doi:10.1007/3-540-48405-1\_14}}.

\bibitem{Broder1997}
Andrei~Z. Broder.
\newblock On the resemblance and containment of documents.
\newblock In {\em Proc. {SEQUENCES}}, pages 21--29, 1997.

\bibitem{BCFM2000}
Andrei~Z. Broder, Moses Charikar, Alan~M. Frieze, and Michael Mitzenmacher.
\newblock Min-wise independent permutations.
\newblock {\em J. Comput. Syst. Sci.}, 60(3):630--659, 2000.
\newblock \href {https://doi.org/10.1006/jcss.1999.1690}
  {\path{doi:10.1006/jcss.1999.1690}}.

\bibitem{BTC2006}
Benjamin~C. Brodie, David~E. Taylor, and Ron~K. Cytron.
\newblock A scalable architecture for high-throughput regular-expression
  pattern matching.
\newblock In {\em Proc. 33rd {ISCA}}, pages 191--202, 2006.
\newblock \href {https://doi.org/10.1109/ISCA.2006.7}
  {\path{doi:10.1109/ISCA.2006.7}}.

\bibitem{Charikar2002}
Moses Charikar.
\newblock Similarity estimation techniques from rounding algorithms.
\newblock In {\em Proc. 34th STOC}, pages 380--388, 2002.
\newblock \href {https://doi.org/10.1145/509907.509965}
  {\path{doi:10.1145/509907.509965}}.

\bibitem{DAS2010}
Shuai Ding, Josh Attenberg, and Torsten Suel.
\newblock Scalable techniques for document identifier assignment in inverted
  indexes.
\newblock In {\em Proc. 19th {WWW}}, pages 311--320, 2010.

\bibitem{DI2003}
Fred Douglis and Arun Iyengar.
\newblock Application-specific delta-encoding via resemblance detection.
\newblock In {\em Proc. {USENIX} ATC, General Track 2003}, pages 113--126,
  2003.
\newblock URL: \url{http://www.usenix.org/events/usenix03/tech/douglis.html}.

\bibitem{FHP1981}
Arthur~M. Farley, Stephen~T. Hedetniemi, and Andrzej Proskurowski.
\newblock Partitioning trees: Matching, domination, and maximum diameter.
\newblock {\em Int. J. Parallel Program.}, 10(1):55--61, 1981.
\newblock \href {https://doi.org/10.1007/BF00978378}
  {\path{doi:10.1007/BF00978378}}.

\bibitem{FGPV+2008}
Domenico Ficara, Stefano Giordano, Gregorio Procissi, Fabio Vitucci, Gianni
  Antichi, and Andrea~Di Pietro.
\newblock An improved {DFA} for fast regular expression matching.
\newblock {\em Comput. Commun. Rev.}, 38(5):29--40, 2008.
\newblock \href {https://doi.org/10.1145/1452335.1452339}
  {\path{doi:10.1145/1452335.1452339}}.

\bibitem{FPGP+2011}
Domenico Ficara, Andrea~Di Pietro, Stefano Giordano, Gregorio Procissi, Fabio
  Vitucci, and Gianni Antichi.
\newblock Differential encoding of dfas for fast regular expression matching.
\newblock {\em {IEEE/ACM} Trans. Netw.}, 19(3):683--694, 2011.
\newblock \href {https://doi.org/10.1109/TNET.2010.2089639}
  {\path{doi:10.1109/TNET.2010.2089639}}.

\bibitem{GWXC+2023}
Lei Gong, Chao Wang, Haojun Xia, Xianglan Chen, Xi~Li, and Xuehai Zhou.
\newblock Enabling fast and memory-efficient acceleration for pattern matching
  workloads: The lightweight automata processing engine.
\newblock {\em {IEEE} Trans. Computers}, 72(4):1011--1025, 2023.
\newblock \href {https://doi.org/10.1109/TC.2022.3187338}
  {\path{doi:10.1109/TC.2022.3187338}}.

\bibitem{PIM2012}
Sariel Har{-}Peled, Piotr Indyk, and Rajeev Motwani.
\newblock Approximate nearest neighbor: Towards removing the curse of
  dimensionality.
\newblock {\em Theory Comput.}, 8(1):321--350, 2012.
\newblock \href {https://doi.org/10.4086/toc.2012.v008a014}
  {\path{doi:10.4086/toc.2012.v008a014}}.

\bibitem{BM2020}
Martin Hemmingsen and Benjamin~Wrist Lam.
\newblock Fast compression of {DFA}s for intrusion detection systems.
\newblock Master's thesis, Tech. Uni. Denmark., 2021.

\bibitem{KSE2008}
Shijin Kong, Randy Smith, and Cristian Estan.
\newblock Efficient signature matching with multiple alphabet compression
  tables.
\newblock In {\em Proc. 4th {SECURECOMM}}, page~1, 2008.
\newblock \href {https://doi.org/10.1145/1460877.1460879}
  {\path{doi:10.1145/1460877.1460879}}.

\bibitem{KH2015}
Lubos Krc{\'{a}}l and Jan Holub.
\newblock Incremental locality and clustering-based compression.
\newblock In {\em {DCC} 2015}, pages 203--212, 2015.
\newblock \href {https://doi.org/10.1109/DCC.2015.23}
  {\path{doi:10.1109/DCC.2015.23}}.

\bibitem{Kruskal1956}
Joseph~B Kruskal.
\newblock On the shortest spanning subtree of a graph and the traveling
  salesman problem.
\newblock {\em Proc. Amer. Math. Soc.}, 7(1):48--50, 1956.

\bibitem{KDLT2004}
Purushottam Kulkarni, Fred Douglis, Jason~D. LaVoie, and John~M. Tracey.
\newblock Redundancy elimination within large collections of files.
\newblock In {\em Proc. {USENIX} ATC, General Track 2004}, pages 59--72, 2004.

\bibitem{KDYP+2006}
Sailesh Kumar, Sarang Dharmapurikar, Fang Yu, Patrick Crowley, and Jonathan~S.
  Turner.
\newblock Algorithms to accelerate multiple regular expressions matching for
  deep packet inspection.
\newblock In {\em Proc. SIGCOMM 2006}, pages 339--350, 2006.
\newblock \href {https://doi.org/10.1145/1159913.1159952}
  {\path{doi:10.1145/1159913.1159952}}.

\bibitem{KTW2006}
Sailesh Kumar, Jonathan~S. Turner, and John Williams.
\newblock Advanced algorithms for fast and scalable deep packet inspection.
\newblock In {\em Proc. {ANCS} 2006}, pages 81--92, 2006.
\newblock \href {https://doi.org/10.1145/1185347.1185359}
  {\path{doi:10.1145/1185347.1185359}}.

\bibitem{LT2014}
Alex~X. Liu and Eric Torng.
\newblock An overlay automata approach to regular expression matching.
\newblock In {\em Proc. 33rd {INFOCOM}}, pages 952--960, 2014.
\newblock \href {https://doi.org/10.1109/INFOCOM.2014.6848024}
  {\path{doi:10.1109/INFOCOM.2014.6848024}}.

\bibitem{LSLH+2017}
Shu Liu, Shaojing Su, Desheng Liu, Zhiping Huang, and Mingyan Xiao.
\newblock Efficient compression algorithm for ternary content addressable
  memory-based regular expression matching.
\newblock {\em Electronics Letters}, 53(3):152--154, 2017.

\bibitem{MKMK2018}
Denis Matousek, Juraj Kubis, Jir{\'{\i}} Matousek, and Jan Korenek.
\newblock Regular expression matching with pipelined delayed input dfas for
  high-speed networks.
\newblock In {\em Proc. {ANCS} 2018}, pages 104--110, 2018.
\newblock \href {https://doi.org/10.1145/3230718.3230730}
  {\path{doi:10.1145/3230718.3230730}}.

\bibitem{MMK2018}
Denis Matousek, Jir{\'{\i}} Matousek, and Jan Korenek.
\newblock High-speed regular expression matching with pipelined memory-based
  automata.
\newblock In {\em Proc. 26th {FCCM}}, page 214, 2018.
\newblock \href {https://doi.org/10.1109/FCCM.2018.00048}
  {\path{doi:10.1109/FCCM.2018.00048}}.

\bibitem{MPNT+2010}
Chad~R. Meiners, Jignesh Patel, Eric Norige, Eric Torng, and Alex~X. Liu.
\newblock Fast regular expression matching using small tcams for network
  intrusion detection and prevention systems.
\newblock In {\em 19th {USENIX} Security}, pages 111--126, 2010.

\bibitem{OMST2002}
Zan Ouyang, Nasir~D. Memon, Torsten Suel, and Dimitre Trendafilov.
\newblock Cluster-based delta compression of a collection of files.
\newblock In {\em Proc. 3rd {WISE}}, pages 257--268, 2002.
\newblock \href {https://doi.org/10.1109/WISE.2002.1181662}
  {\path{doi:10.1109/WISE.2002.1181662}}.

\bibitem{PLT2014}
Jignesh Patel, Alex~X. Liu, and Eric Torng.
\newblock Bypassing space explosion in high-speed regular expression matching.
\newblock {\em {IEEE/ACM} Trans. Netw.}, 22(6):1701--1714, 2014.
\newblock \href {https://doi.org/10.1109/TNET.2014.2309014}
  {\path{doi:10.1109/TNET.2014.2309014}}.

\bibitem{Paxson1999}
Vern Paxson.
\newblock Bro: a system for detecting network intruders in real-time.
\newblock {\em Comput. Networks}, 31(23-24):2435--2463, 1999.

\bibitem{PWZ2011}
Andrew Peel, Anthony Wirth, and Justin Zobel.
\newblock Collection-based compression using discovered long matching strings.
\newblock In {\em Proc. 20th {CIKM}}, pages 2361--2364, 2011.
\newblock \href {https://doi.org/10.1145/2063576.2063967}
  {\path{doi:10.1145/2063576.2063967}}.

\bibitem{Prim1957}
Robert~Clay Prim.
\newblock Shortest connection networks and some generalizations.
\newblock {\em The Bell System Technical Journal}, 36(6):1389--1401, 1957.

\bibitem{SS2017}
S~Prithi and S~Sumathi.
\newblock A survey on recent dfa compression techniques for deep packet
  inspection in network intrusion detection system.
\newblock {\em Journal of Electrical Engineering}, 17(3):14--14, 2017.

\bibitem{QWFX+2011}
Yaxuan Qi, Kai Wang, Jeffrey Fong, Yibo Xue, Jun Li, Weirong Jiang, and
  Viktor~K. Prasanna.
\newblock {FEACAN:} front-end acceleration for content-aware network
  processing.
\newblock In {\em Proc. 30th {INFOCOM}}, pages 2114--2122, 2011.
\newblock \href {https://doi.org/10.1109/INFCOM.2011.5935021}
  {\path{doi:10.1109/INFCOM.2011.5935021}}.

\bibitem{Roesch1999}
Martin Roesch.
\newblock Snort: Lightweight intrusion detection for networks.
\newblock In {\em Proc. 13th {LISA}}, pages 229--238, 1999.

\bibitem{Roussev2010}
Vassil Roussev.
\newblock Data fingerprinting with similarity digests.
\newblock In {\em {IFIP} Int. Conf. Digital Forensics 2010}, volume 337, pages
  207--226, 2010.
\newblock \href {https://doi.org/10.1007/978-3-642-15506-2\_15}
  {\path{doi:10.1007/978-3-642-15506-2\_15}}.

\bibitem{SLHW2017}
Subramanian~Shiva Shankar, Pinxing Lin, Andreas Herkersdorf, and Thomas Wild.
\newblock A divide and conquer state grouping method for bitmap based
  transition compression.
\newblock In {\em Proc. 18th {PDCAT}}, pages 400--406, 2017.
\newblock \href {https://doi.org/10.1109/PDCAT.2017.00071}
  {\path{doi:10.1109/PDCAT.2017.00071}}.

\bibitem{SHWH2012}
Philip Shilane, Mark Huang, Grant Wallace, and Windsor Hsu.
\newblock Wan-optimized replication of backup datasets using stream-informed
  delta compression.
\newblock {\em {ACM} Trans. Storage}, 8(4):13:1--13:26, 2012.
\newblock \href {https://doi.org/10.1145/2385603.2385606}
  {\path{doi:10.1145/2385603.2385606}}.

\bibitem{TJDS+2017}
Qiu Tang, Lei Jiang, Qiong Dai, Majing Su, Hongtao Xie, and Binxing Fang.
\newblock {RICS-DFA:} a space and time-efficient signature matching algorithm
  with reduced input character set.
\newblock {\em Concurr. Comput. Pract. Exp.}, 29(20), 2017.
\newblock \href {https://doi.org/10.1002/cpe.3940}
  {\path{doi:10.1002/cpe.3940}}.

\bibitem{TSCV2004}
Nathan Tuck, Timothy Sherwood, Brad Calder, and George Varghese.
\newblock Deterministic memory-efficient string matching algorithms for
  intrusion detection.
\newblock In {\em Proc. 23rd {INFOCOM}}, pages 2628--2639, 2004.
\newblock \href {https://doi.org/10.1109/INFCOM.2004.1354682}
  {\path{doi:10.1109/INFCOM.2004.1354682}}.

\bibitem{XJFH2011}
Wen Xia, Hong Jiang, Dan Feng, and Yu~Hua.
\newblock Silo: {A} similarity-locality based near-exact deduplication scheme
  with low {RAM} overhead and high throughput.
\newblock In {\em {USENIX} ATC 2011}, 2011.

\bibitem{XCSY+2016}
Chengcheng Xu, Shuhui Chen, Jinshu Su, Siu{-}Ming Yiu, and Lucas Chi~Kwong Hui.
\newblock A survey on regular expression matching for deep packet inspection:
  Applications, algorithms, and hardware platforms.
\newblock {\em {IEEE} Commun. Surv. Tutorials}, 18(4):2991--3029, 2016.
\newblock \href {https://doi.org/10.1109/COMST.2016.2566669}
  {\path{doi:10.1109/COMST.2016.2566669}}.

\bibitem{YCDL+2006}
Fang Yu, Zhifeng Chen, Yanlei Diao, T.~V. Lakshman, and Randy~H. Katz.
\newblock Fast and memory-efficient regular expression matching for deep packet
  inspection.
\newblock In {\em Proc. {ANCS} 2006}, pages 93--102, 2006.
\newblock \href {https://doi.org/10.1145/1185347.1185360}
  {\path{doi:10.1145/1185347.1185360}}.

\end{thebibliography}

\appendix
\section{Input DFA Characteristics}
\begin{table}[h]
    \centering
    \begin{tabularx}{\textwidth}{lXXXXXXX}
      Source & States & Rules & Average length of rules & \% using wildcards (\texttt{*}, \texttt{+}, \texttt{?}) & \% using length restrictions (\texttt{\{,k,+\}}) \\ \hline
        Snort & 991 & 12 & 34.3 & 50.0 & 33.3 \\
        Snort & 1691 & 14 & 35.2 & 57.1 & 42.9 \\
        Snort & 5667 & 15 & 37.3 & 60.0 & 46.7 \\
        Snort & 8467 & 22 & 35.3 & 59.1 & 40.9 \\
        Snort & 15902 & 50 & 29.3 & 46.0 & 30.0 \\
        Snort & 39244 & 61 & 28.5 & 41.0 & 24.6 \\
        Snort & 83055 & 101 & 28.0 & 53.5 & 15.8 \\
        Snort & 144413 & 102 & 28.3 & 53.9 & 16.7 \\
        Snort & 246547 & 128 & 29.9 & 49.2 & 23.4 \\
        Snort & 523737 & 134 & 30.0 & 48.5 & 25.4 \\
        Snort & 1069239 & 175 & 37.9 & 55.4 & 39.4 \\
        Suricata & 1066 & 25 & 101.6 & 92.0 & 0.0 \\
        Suricata & 2691 & 28 & 106.2 & 89.3 & 0.0 \\
        Suricata & 6211 & 30 & 112.8 & 90.0 & 0.0 \\
        Suricata & 13577 & 32 & 117.2 & 90.6 & 0.0 \\
        Suricata & 32014 & 38 & 111.1 & 86.8 & 2.6 \\
        Suricata & 75596 & 49 & 105.7 & 85.7 & 2.0 \\
        Suricata & 128696 & 93 & 103.6 & 87.1 & 1.1 \\
        Suricata & 273904 & 101 & 103.0 & 87.1 & 1.0 \\
        Suricata & 462248 & 180 & 101.0 & 89.4 & 0.6 \\
        Suricata & 995152 & 194 & 99.1 & 87.1 & 0.5 \\
        Zeek & 2338 & 5 & 33.2 & 60.0 & 40.0 \\
        Zeek & 4265 & 13 & 46.0 & 61.5 & 15.4 \\
        Zeek & 10137 & 14 & 46.2 & 64.3 & 14.3 \\
        Zeek & 12747 & 17 & 40.7 & 52.9 & 11.8 \\
        Zeek & 30131 & 18 & 42.3 & 55.6 & 11.1 \\
        Zeek & 67933 & 19 & 43.7 & 57.9 & 10.5 \\
        Zeek & 127839 & 20 & 43.6 & 60.0 & 10.0 \\
        Zeek & 340279 & 21 & 44.6 & 61.9 & 9.5 \\
        Zeek & 384949 & 27 & 39.6 & 55.6 & 7.4 \\
        Zeek & 1036954 & 29 & 39.6 & 58.6 & 10.3 \\
    \end{tabularx}
    \caption{DFAs used for testing and characteristics of the corresponding set of regular expressions.}
    \label{tab:rule_characteristics}
\end{table}

\end{document}